\documentclass[structabstract]{aa}
\usepackage{amsmath}
\usepackage{txfonts}
\usepackage{graphicx}
\usepackage{natbib}
\bibpunct{(}{)}{;}{a}{}{,}  

\def\dmu{\mbox{pc cm$^{-3}$}}

\title{Strong pulses detected from a rotating radio transient J1819$-$1458}

\author{H. D. Hu\inst{1,2}
\and A. Esamdin\inst{1}\thanks{Email: aliyi@uao.ac.cn}
\and J. P. Yuan\inst{1}
\and Z. Y. Liu\inst{1}
\and R. X. Xu\inst{3}
\and J. Li\inst{1,2}
\and G. C. Tao\inst{1,2}
\and N. Wang\inst{1}
}

\institute{Urumqi Observatory, NAOC, 40-5 South Beijing Road, Urumqi 830011, P. R. China
\and Graduate University of Chinese Academy of Sciences, 19A Yuquan Road, Beijing 100049, P. R. China
\and School of Physics and State Key Laboratory of Nuclear Physics and Technology, Peking University, Beijing 100871, P. R. China
}
\begin{document}
\date{\today}
\abstract
{}
{We analyze individual pulses
detected from RRAT J1819$-$1458.}
{From April 2007 to April 2010,
we carried out observations
using the Nanshan 25-m radio telescope
of Urumqi Observatory
at a central frequency of 1541.25 MHz.}
{We obtain a dispersion measure
$DM=195.7\pm0.3$ \dmu~by analyzing
all the 423 detected bursts.
The tri-band pattern
of arrival time residuals
is confirmed
by a single pulse timing analysis.
Twenty-seven bimodal bursts
located in the middle residual band
are detected,
and, profiles of
two typical bimodal bursts and
two individual single-peak pulses
are presented.
We determine the statistical properties
of SNR and W$_{50}$
of bursts
in different residual bands.
The W$_{50}$ variation
with SNR shows that
the shapes of bursts
are quite different
from each other.
The cumulative probability distribution
of intensity
for a possible power law with
index $\alpha=1.6\pm0.2$ is inferred
from the number of
those bursts with $SNR\ge6$
and high intensities.}
{}

\maketitle

\section{Introduction}\label{sec_intro}
Rotating radio transients (RRATs)
are a recently discovered class of neutron stars.
At a wave band $\sim$1.4 GHz,
they have sporadic radio pulses
of durations ranging from 0.5 to 100 ms
with flux densities
from $\sim$10 mJy to $\sim$10 Jy
and burst rates
from 0.3 to 50.3 per hour.
The study of the time of arrival (TOA)
of single pulses
indicates that these RRATs have periods
ranging from 0.1 to 7 s
and surface magnetic field
of $\sim$$10^{12}$ or $\sim$$10^{13}$ gauss
\citep{mll+06,dcm+2009,mlk+09,bb2010,kle+2010}.

J1819$-$1458 is the brightest
and most prolific
radio transient
among the more than forty known RRATs.
The dispersion measure of the source is
$DM\approx196$ \dmu.
At 1.4 GHz,
the durations of most bursts are
$\sim$3 ms at intervals of
$\sim$3 minutes
and the peak flux density
is $\sim$10 Jy \citep{mll+06,esamdin+08,lmk+2009}.
The irregular bursts cannot be detected
by standard periodicity search methods,
but a single pulse TOA analysis
is capable of identifying
the rotation period of $P\approx4.26$ s
with a spin-down rate
$\dot{P}\approx5.76\times 10^{-13}$ \mbox{s s$^{-1}$}.
The inferred surface magnetic field density
is $B_\text s\approx5\times10^{13}$ gauss
and characteristic age
is $\tau_\text c\approx1.2\times10^5$ years
\citep{mll+06,esamdin+08}.
\citet{lmk+2009} reported two unusual glitches
with quite different post-glitch behavior
and an associated increase in the flux intensity,
which implies
that RRAT J1819$-$1458
may be moving toward
the death line
in the pulsar $P-\dot P$ diagram.
The Faraday rotation measure
of RRAT J1819$-$1458 is
$RM\approx330\pm30$ \mbox{rad m$^{-2}$}
and the integrated linear polarization measurement
is relatively low
owing to orthogonal modes of polarization,
although some individual pulses are highly polarized
\citep{khs+2009}.

After the {\em{Chandra~X-ray~Observatory}}
detected the X-ray counterpart
\citep{rbg+2006,gmr+2007},
the {\em{XMM-Newton}} discovered
the X-ray pulsation aligned
with the phase of the radio pulse and
a possible X-ray spectral feature that is similar
to those of the X-ray dim isolated neutron stars
\citep[XDINS, cf.][]{ptp06,hab07,kml+2009}.
This indicates that
RRAT J1819$-$1458 is a cooling neutron star and
possibly a transitional object between the pulsar
and magnetar classes (\citealt[][]{mrg+07};
for descriptions of magnetars,
see \citealt{wt06,mer2008}).
\citet{rmg+2009} found the X-ray extended emission
around RRAT J1819$-$1458,
which can be interpreted as being
a nebula powered by the pulsar.
The near-infrared counterpart
to the extended X-ray emission
was not detected
in the observation taken by \citet{rct+2010}.
There is no evidence of the optical counterpart
to the RRAT in the observation implemented by
\citet{dml2006},
but there is possibly
a near-infrared counterpart \citep{rct+2010}.

\citet{wsr+2006}
assumed that some RRATs could be explained
in terms of normal pulsars
that have pulse amplitude modulations,
such as PSR B0656+14.
Had they been put at properly large distances,
their weak pulses would have been undetectable
and they might have been identified as RRATs.
From the observation results
of PSR J1119$-$6127
with magnetic field strength
and post-glitch behavior
similar to those of
RRAT J1819$-$1458,
it is understood
that some RRATs can be interpreted
as the consequences of
the line of sight
missing the steady pulse components
of emission area
and intercepting these unstable ones
\citep{wje2010}.
\citet{wmj07} proposed
that RRATs could not be directly
related to nulling pulsars
because the latter
have no observed pulses
such as giant pulses
\citep[e.g.][]{hkw+03,kbm+06}.
However,
taking PSR J0941$-$39
as an example,
\citet{bb2010} argued
that there was possibly
a link between RRATs
and pulsars
with nulling phenomena.
There is also a hypothesis that
some RRATs are possibly related to
radio pulsing magnetars such as XTE J1810$-$197
\citep{ssw+2009}.
By studying the birthrates of neutron stars,
RRATs have been considered distant XDINS \citep{ptp06}
or an evolutionary stage
of a single class of objects \citep{kk2008}.
In the $P-\dot P$ diagram,
the periods of known RRATs
cover a large range
where periods
of normal pulsars
and magnetars
distribute;
the observational characteristics of RRATs vary
with diversities \citep{mlk+09}.
Although there have been several theoretical interpretations
of RRATs \citep[e.g.][]{li06,zgd07,lm07,cs2008,olny2009},
RRATs do not appear to be completely understood.

This paper presents a study of individual bursts detected
in single pulse observations from April 2007 to April 2010.
Section \ref{sec_obs} describes the observation details.
The results of study on detected pulses
are given in Sect.\ref{sec_res}
and discussed in Sect.\ref{sec_dis}.
Section \ref{sec_con} briefly summarizes this paper.

\section{Observations}\label{sec_obs}
Observations were performed
using the Nanshan 25-m radio telescope of Urumqi Observatory.
The antenna has a cryogenic receiver
of two orthogonal linear polarization channels
with a central frequency of 1541.25 MHz.
Each polarization channel
consists of
128 sub-channels
of bandwidth 2.5 MHz
and has a total bandwidth of 320 MHz.
The signal from each sub-channel is 1-bit sampled
and recorded to hard disk for subsequent off-line processing
\citep[for more details of the system, see][]{wmz+01}.
The minimum detectable flux density
$S_\text{min}\approx3.4$ Jy is limited by
\begin{equation}\label{equ_Smin}
S_\text{min}=\frac{2\alpha\beta{kT_\text{sys}}}{\eta{A\sqrt{n_\text p\tau{\Delta{f}}}}},
\end{equation}
in which $\alpha=5$ is the threshold signal-to-noise ratio (SNR),
$\beta=\sqrt{\pi/2}$ is the loss factor due to 1-bit digitization,
$k$ is the Boltzmann constant,
$T_\text{sys}=T_\text{rec}+T_\text{spl}+T_\text{sky}\approx32$ K is the system temperature,
$\eta\approx57\%$ is the telescope efficiency at 1541.25 MHz,
$A=490.9$ m$^2$ is the telescope area,
$n_\text p=2$ is the number of polarization channels,
$\tau=0.5$ ms is the sampling interval,
and $\Delta{f}=320$ MHz is the total bandwidth \citep{mlc+2001}.

From Apr 10, 2007
to Apr 13, 2010,
260 hours of data were accumulated
in 1100 days of observations.
The time-span of each observation session
was 2 hours.

\section{Data analysis and results}\label{sec_res}
\begin{figure}
\resizebox{\hsize}{!}{\includegraphics[angle=270]{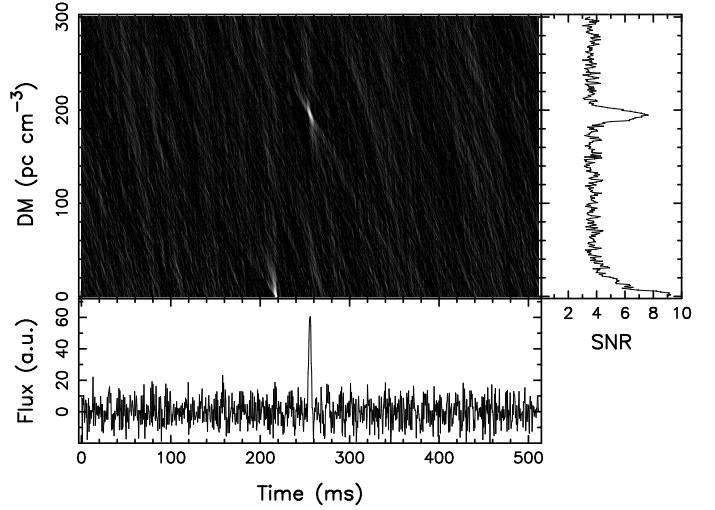}}
\caption{Diagnostic plot showing detected burst
from RRAT J1819$-$1458 and RFI signal.
The upper-left panel is a DM-time gray map generated by
de-dispersing the signal with $DM$ varying
from 0.7 to 300.7 \dmu~in steps of 1 \dmu.
The brightness of the plot
indicates the amplitude
at the corresponding position,
and, the two bright marks at around
$DM=195.7$ and $DM=0$
\dmu~represent the burst and RFI.
The upper-right panel is a DM--SNR graph
with two peak SNR values at around $DM=195.7$ and $DM=0$ \dmu,
in which the SNR for each $DM$ is calculated
from the bin of maximum amplitude over 512 ms.
The lower-left panel is a time series graph
showing the amplitude (in an arbitrary unit)
of each bin at $DM=195.7$ \dmu,
which clearly delineates a narrow burst profile at the center.}
\label{fig_gray}
\end{figure}

Pulsed radio waves propagating through
the ionized plasma of the interstellar media
are dispersed because
the group velocity in the media
is frequency dependent.
This demonstrates
that the higher frequency components
of a radio pulse arrive
before those at lower frequencies.
To achieve a higher SNR and narrower
(normally closer to the real) profile of the pulse,
the time delay must be removed.
The time difference between the two components
of different frequencies, $\Delta{t}$,
is given by
\begin{equation}\label{equ_dt}
\Delta{t}=4.1488\times{DM}\times(\frac{1}{f^2_\text l}-\frac{1}{f^2_\text h})\textrm{ ms,}
\end{equation}
where $DM$ is the dispersion measure in \dmu,
and $f_\text l$ and $f_\text h$ are the values (in GHz)
of the lower and higher frequencies, respectively
\citep[][Appendix 2.4]{lk05}.
We use Eq.\ref{equ_dt}
with the nominal DM value of
$DM=196$ \dmu~to calculate
and remove the delayed time of signal
between the sub-channel at the lowest frequency
and sub-channels at other frequencies,
finding that the signals
from two polarization channels
are de-dispersed.
The signals of all 256 sub-channels are summed,
and signals of $SNR\ge5$ are searched for
in the whole time series of each observation.
Selected signals
with SNR above the threshold
are reprocessed with $DM$
varying from 0.7 to 300.7 \dmu~in steps
of 1 \dmu.
To each $DM$,
the SNR of the bin with the maximum amplitude is
computed over 1024 bins (512 ms),
with the candidate burst being at the center of the series.
Through the process,
a diagnostic plot shown in
Fig.\ref{fig_gray}
is generated and used for careful inspection
by eye to identify real bursts
\citep[for more details of the burst detection method adopted, see][]{esamdin+08}.

In the upper-left panel
of Fig.\ref{fig_gray},
we present a DM-time gray map
that obviously shows two white marks.
The mark at around
the intersection point
of the line of $DM=195.7$
\dmu~and center line of the time series
represents a burst detected
from RRAT J1819$-$1458,
and the one at about 210 ms around $DM=0$
\dmu~represents a signal of
radio frequency interference
\citep[RFI, see][]{cm03}
that sometimes contaminates
the observation data.
The mark representing the burst dissolves
as the $DM$ increases or decreases
from the nominal $DM$,
whereas the one representing the RFI dissolves
as the $DM$ increases from 0 \dmu.
The upper-right panel is a DM--SNR line graph
illustrating the SNR variation with DM,
where the SNR is
from the bin of maximum amplitude
in the 512 ms time series
for every DM value.
As illustrated,
the SNR has two peak values
at around $DM=195.7$ \dmu~for the burst
and $DM=0$ \dmu~for the RFI,
and far away from these two DM values,
the SNR varies like white noise.
The lower-left panel shows the time series
at $DM=195.7$
\dmu~and clearly displays a burst profile in the center.
A typical burst detected
from RRAT J1819$-$1458 has
the visual characteristics
presented in Fig.\ref{fig_gray},
which is a significant criterion
for identifying a real burst.

Using the above method,
423 strong pulses
of SNR ranging from 5 to 13.3 and
durations from 0.5 to 13.7 ms are detected.
As \citet{esamdin+08} noted
that five bimodal bursts were detected,
there are 27 bimodal bursts detected
with distances between two component peaks
ranging from 3 to 15 ms,
SNR from 5.5 to 9.5,
and full widths at half maximum (FWHM, W$_{50}$)
from 1.3 to 7.2 ms.
For a bimodal burst,
the SNR means the higher SNR of the two peaks
and the W$_{50}$ is calculated
from the peak of the higher SNR.

The W$_{50}$ of a burst is obtained from
\begin{equation}\label{equ_w50}
W_{50}=\sqrt{W_\text{50,obs}^2-t_\text{samp}^2-t_\text{DM}^2-t_\text{scatt}^2},
\end{equation}
in which $W_\text{50,obs}$ is
the observed FWHM,
$t_\text{samp}=0.5$ ms is the sampling interval,
$t_\text{DM}=1.109$ ms is the dispersion smearing time
across one sub-channel,
and $t_\text{scatt}=0.015$ ms is the broadening of the pulse
due to interstellar scattering.
Both $t_\text{DM}$ and $t_\text{scatt}$ are given by
\begin{equation}\label{equ_tdm}
t_\text{DM}=8.2976\times10^6\times{DM}\times{f_\text c^{-3}}\times{B}\textrm{ ms}
\end{equation}
and
\begin{equation}\label{equ_tscat}
t_\text{scatt}=(\frac{DM}{1000})^{3.5}\times(\frac{400}{f_\text c})^4\times{10^3}\textrm{ ms,}
\end{equation}
where $DM=195.7$ \dmu~is the dispersion measure,
$f_\text c=1541.25$ MHz is the central frequency of the receiver channel,
and $B=2.5$ MHz is the sub-channel bandwidth
\citep{lgs06,lk05}.

\begin{figure}
\resizebox{\hsize}{!}{\includegraphics{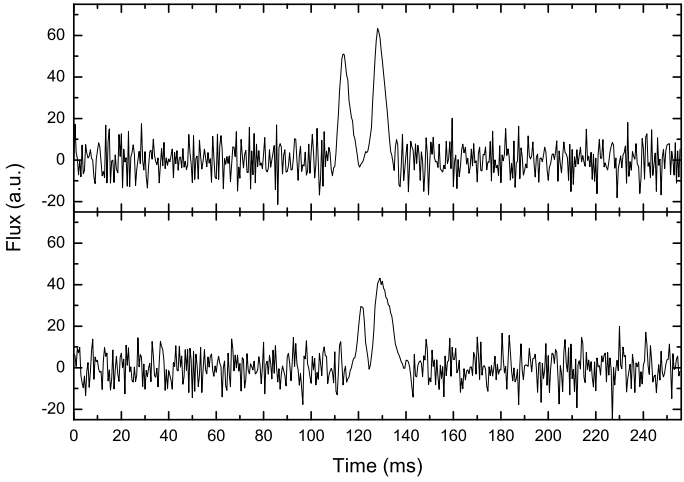}}
\caption{Profiles of two typical bimodal bursts
detected from RRAT J1819$-$1458
where flux density is given in arbitrary unit.
The higher SNR of the two peaks
in each bimodal bursts
are 9.2 and 6.5
in the top and bottom panels, respectively.
Distances between the two peaks
and W$_{50}$ of peaks with the higher SNR
in each bimodal burst
are 14.5 and 4.5 ms in the top panel
and 8 and 7.2 ms in the bottom panel.}
\label{fig_bim_prof}
\end{figure}

Fig.\ref{fig_bim_prof} portrays two profiles
of typical detected bimodal bursts
from RRAT J1819$-$1458
with the distances between their two peaks
being 14.5 and 8 ms.
In Fig.\ref{fig_bim_prof},
the SNR and W$_{50}$
are 9.2 and 4.5 ms in the top panel,
and, 6.5 and 7.2 ms in the bottom panel.
The peaks of the higher SNR
appear behind
those of the lower SNR,
as shown in Fig.\ref{fig_bim_prof},
in 16 bimodal bursts,
while peaks of the lower SNR
appear behind those
of the higher SNR
in another nine bimodal bursts.
In the remaining two bimodal bursts,
the two peaks visibly
have approximately equal SNR.
There are four bimodal bursts
that are possibly ``tri-modal",
which means that one burst
has possibly three close-set peaks
with distances
between two adjacent peaks
less than 5 ms.
However,
it is difficult to distinguish
the third peak from noise fluctuations
because of the low flux density (SNR $\le3$).

\begin{figure}
\resizebox{\hsize}{!}{\includegraphics{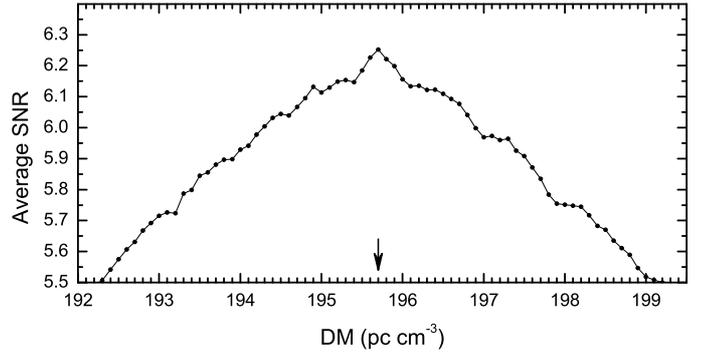}}
\caption{Variation of average SNR with DM.
The average SNR
that has a peak value at $DM=195.7$
\dmu~as indicated by the arrow,
is calculated
from all 423 detected bursts from RRAT J1819$-$1458
with $DM$ increasing
from 192 to 199.5 \dmu~in steps of 0.1 \dmu.}
\label{fig_snr_dm}
\end{figure}

To obtain a $DM$
closer to the intrinsic value,
with $DM$
increasing from 0 to 400
\dmu~in steps of 0.1 \dmu,
average SNR of all 423 bursts
from RRAT J1819$-$1458
are calculated
for all $DM$.
Fig.\ref{fig_snr_dm} is the graph
of average SNR versus DM
with $DM$ varying from 192 to 199.5
\dmu~and shows that
the peak SNR is
at $DM=195.7$ \dmu.
In the range of
$DM$ from 300 to 400 \dmu,
where the SNR variation behaves
like white noise,
the standard deviation in the SNR fluctuation
is 0.023 and its quintuple is 0.115.
The $DM$ uncertainty
is then limited to 0.3
\dmu~at a 5$\sigma$ confidence level.
The result $DM=195.7\pm0.3$
\dmu~is in accord with those of
\citet{mll+06} and \citet{esamdin+08}.

\begin{table}
\centering
    \caption{\label{tab_tim_para}
    Timing parameters of RRAT J1819$-$1458.}
    \begin{tabular}{lcc}
    \hline\hline
    Parameters                                      &   &Values                                 \\
    \hline
    RA (J2000)                                      &   &18:19:34.173                           \\
    Dec (J2000)                                     &   &$-$14:58:03.57                         \\
    Frequency $\nu$ (Hz)                            &   &0.234565100209(7)                      \\
    Frequency derivative $\dot{\nu}$ (Hz s$^{-1}$)  &   &$-3.09123(2)\times{10^{-14}}$          \\
    Epoch (MJD)                                     &   &54321                                  \\
    Dispersion measure $DM$ (\dmu)                  &   &195.7(3)                               \\
    Time span (MJD)                                 &   &54200.1--55299.0                       \\
    Residual rms $\sigma$ (ms)                      &   &7                                      \\
    Fundamental frequency {\sc wave\_om}            &   &0.00437                                \\
    Number of sinusoids $n_\text H$                 &   &13                                     \\
    \hline\hline
    \end{tabular}
    \tablefoot{Numbers in parentheses
    are uncertainties in the last digits,
    which are ten times the {\sc{TEMPO2}} standard errors.
    The position of J1819$-$1458
    is from \citet{rmg+2009},
    of which errors are 0.28 arcsec in both coordinates.
    With TOAs in the bottom and top residual bands
    increased and decreased by 45 ms, respectively,
    $\sigma$ is calculated
    by fitting the parameters
    using the {\sc{TEMPO2}} parameter
    \mbox{\sc{wave\_om}}
    to whiten the timing residuals.}
\end{table}

\begin{figure}
\resizebox{\hsize}{!}{\includegraphics{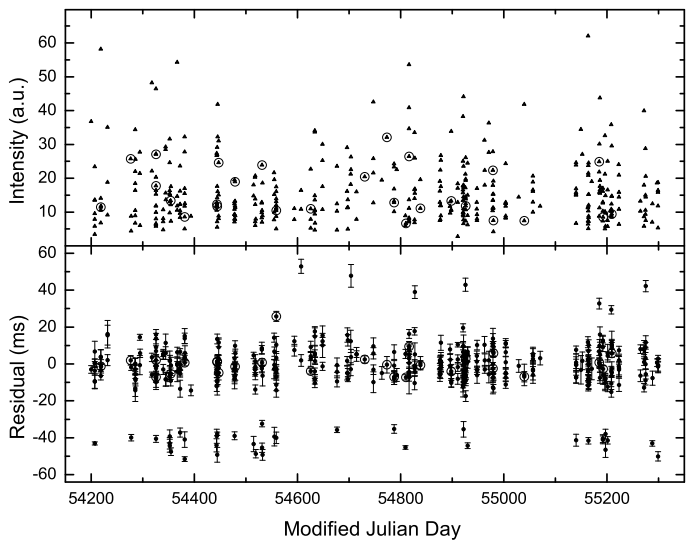}}
\caption{Variation in
flux intensity with date (top)
and timing residuals (bottom)
for RRAT J1819$-$1458.
The flux intensity
is measured in an arbitrary unit
and no remarkable deviation
of flux density
is shown in the variation plot.
The timing residuals
that show a distribution of three bands
with $\sim$40 ms intervals
are calculated with the fitted parameters
listed in Table \ref{tab_tim_para}.
The rms of residuals is 17 ms and
the error bars are 10$\sigma$.
Dots with circles in the graphs
are the counterparts of bimodal bursts
and the gaps in date
are due to observation scheduling.}
\label{fig_int_res}
\end{figure}

In this work,
the flux intensity of a detected burst, $E$,
is quantified by
\begin{equation}\label{equ_int}
E=W_{50}\times{S},
\end{equation}
in which $W_{50}$ is the FWHM
and $S$ the peak flux density.
The epochs of the two glitches
associated with increases
in flux intensity
are beyond
the time range of
this observation
\citep{lmk+2009}.
Thus it is not inconsistent that
there is no burst with
a significant deviation
in flux intensity
as shown
in Fig.\ref{fig_int_res} (top).
From the figure,
there is no evidence
that the intensity of bimodal bursts
undergoes obvious deviation.

As reported by \citet{lmk+2009}
and shown in Fig.\ref{fig_int_res} (bottom),
the timing residuals of RRAT J1819$-$1458
fall into a tri-band distribution.
With {\sc TEMPO2}\footnote{http://www.atnf.csiro.au/research/pulsar/tempo2/}
in {\sc{TEMPO1}} compatibility mode
and using the same method as \citet{lmk+2009},
the parameters $\nu$ and $\dot\nu$
in Table \ref{tab_tim_para}
are obtained
by fitting the 423 TOAs
while increasing and decreasing
the TOAs
in the top and bottom bands
by 45 ms,
respectively.
Taking the two glitches
and different timing epochs into account,
the fitted values
of $\nu$ and $\dot\nu$
correspond to
those acquired by \citet{lmk+2009}.
The residuals from
fitting only $\nu$ and $\dot\nu$
have a waveform
of amplitude $\sim$30 ms
and period of $\sim$2 years.
When the residuals
are calculated
using the unmodified TOAs
and the same fitted parameters,
the residuals
in the two side bands
vary along the tracks
of invariable distances
from the center band;
the residual patterns
for the three bands
are then similar.
This shows that
the residual wave
is not caused by
random phases of the bursts
but probably
by a timing noise
of unknown origin.
The small residual rms
($\sigma=7$ ms)
in Table \ref{tab_tim_para}
is due to
the modification of TOAs
in the top and bottom bands
and using the {\sc{TEMPO2}} parameter
{\sc{wave\_om}}
to remove red noise from the residuals.
The value of {\sc{wave\_om}}
is the fundamental frequency
of the removed wave
and $n_\text H$ is
the number
of the harmonically related sinusoids
\citep[for details of the whitening method, see][]{hem2006}.
The timing residuals
presented in Fig.\ref{fig_int_res} (bottom)
and the rms $\sigma=17$ ms
are computed
from the unmodified TOAs
with parameters that take account of
the whitening parameters
in Table \ref{tab_tim_para}.
No successive bursts are
simultaneously detected
in more than one band
over intervals shorter than
the spin period
and longer
than the largest distance
of two peaks
of the bimodal bursts,
which means that
no more than one burst
is detected in one rotation cycle
at this observation sensitivity.
The middle emission area is more active
and emits about 90\% of the bursts
detected from RRAT J1819$-1458$
including the 27 bimodal ones.

\begin{figure}
\resizebox{\hsize}{!}{\includegraphics{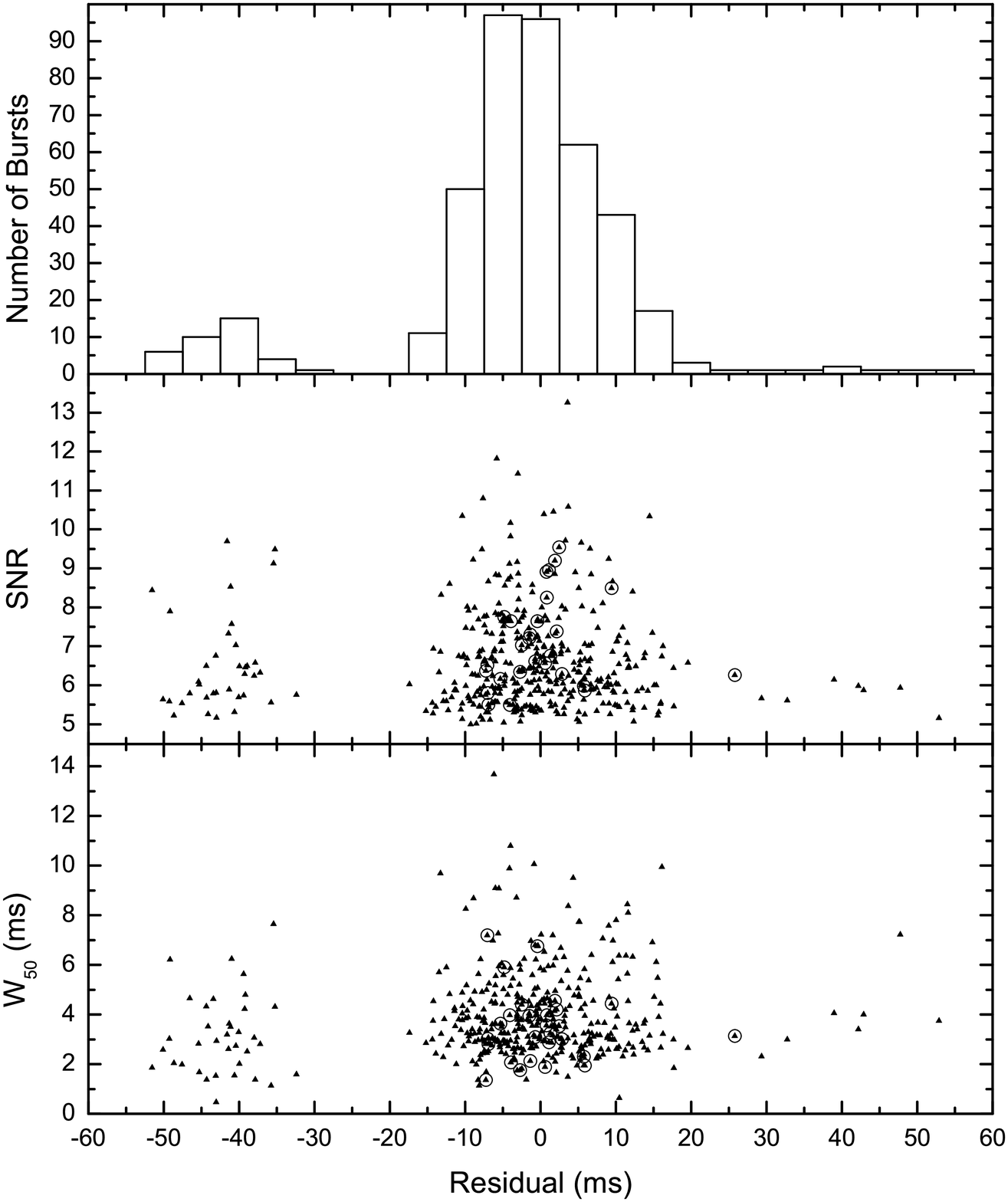}}
\caption{Distribution of timing residuals
of 423 detected bursts
from RRAT J1819$-$1458 (top),
and, plots of SNR (middle)
and W$_{50}$ (bottom) versus timing residual.
Circles in the two plots
are the counterparts of bimodal bursts.}
\label{fig_dist_snr_w50_res}
\end{figure}

Fig.\ref{fig_dist_snr_w50_res} (top)
is the distribution histogram
of the timing residuals
of 423 detected bursts
from RRAT J1819$-$1458,
which shows a tri-modal shape
with only seven bursts scattering
from $\sim$30 to 55 ms
in the last band.
This result agrees with
the residual distribution
reported by \citet{lmk+2009},
considering the high detection threshold
of the flux density
of the Urumqi 25-m telescope
and the low SNR
of bursts in the last band
as shown in Fig.\ref{fig_dist_snr_w50_res} (middle).
The SNR versus timing residual
is plotted in
Fig.\ref{fig_dist_snr_w50_res} (middle),
which shows that
the highest SNR is
from the middle band.
The maximum SNR
of bursts
in the last band
is $\sim$6
and the mean SNR of bursts
in the early band
is 6.5.
It is shown that
SNR of bursts
in the last band
are likely to be lower than
those in the early band.
As displayed
in the bottom panel
of Fig.\ref{fig_dist_snr_w50_res},
the average W$_{50}$ (4.0 ms)
of bursts
in the last band
is slightly larger
than that (3.1 ms)
of bursts
in the early band.
When considering
the poor quality specimen bursts
in the two residual bands,
the mean intensity (14.3 in a.u.)
of bursts in the last band
is approximately equivalent to
that (13.0)
of bursts in the early band,
which is consistent with
the result of
\citet[][Fig.2]{lmk+2009}.

\begin{figure}
\resizebox{\hsize}{!}{\includegraphics{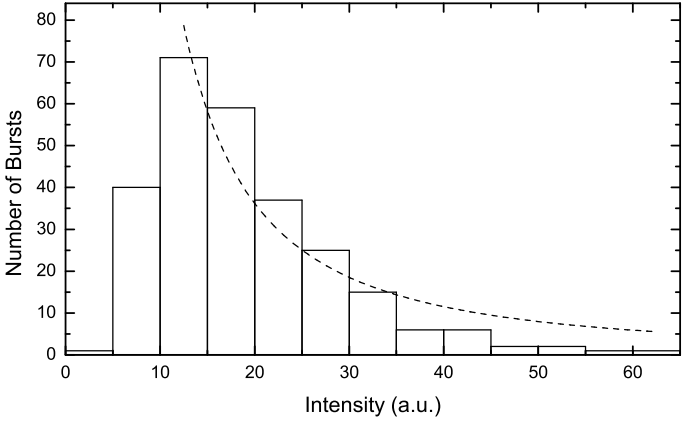}}
\caption{Flux intensity distribution
of bursts detected
from RRAT J1819$-$1458.
The flux intensity distribution
is generated by 266 detected bursts
with $SNR\ge6$.
The dashed curve line
signifies
that the intensity
cumulative probability distribution
is a power law.
The power-law index $\alpha=1.6\pm0.2$
is fitted from
the numbers of bursts with
intensities higher than 10.}
\label{fig_int_dist}
\end{figure}

The histogram in Fig.\ref{fig_int_dist}
is the flux intensity distribution
of 266 detected bursts with $SNR\ge6$
from RRAT J1819$-$1458.
At higher intensities,
the cumulative probability distribution
(CPD) of intensity
is probably a power law.
The power-law index $\alpha=1.6\pm0.2$
with purely formal errors
represents the best fit
for the number of bursts
of intensities higher than 10.
The fitting equation is
\begin{equation}\label{equ_cpd}
CPD=K\times{E^{-\alpha}},
\end{equation}
where $E$ is the flux intensity
in an arbitrary unit \citep{cst+1996}.
Toward both higher and lower intensities,
the deviations
from fitting curve
are possibly
due to
the small amount
of detected bursts
of low intensities
or/and
the CPD actually
not being a power law.
We infer that
it remains unclear whether
the CPD is intrinsically
a power law.

\begin{figure}
\resizebox{\hsize}{!}{\includegraphics{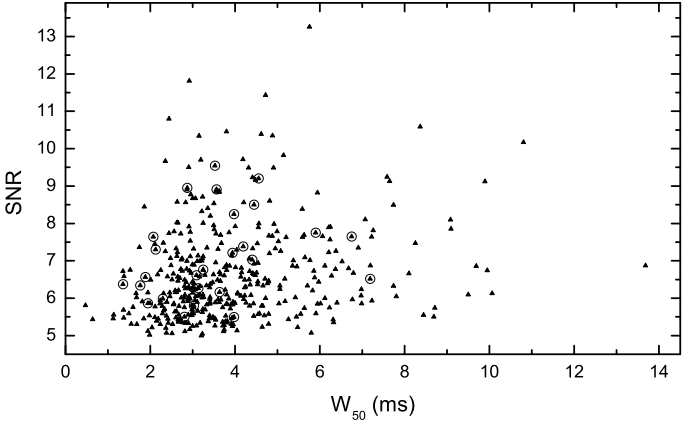}}
\caption{Plot of SNR versus W$_{50}$
of 423 detected bursts from RRAT J1819$-$1458.
The plot displays that
the burst of the highest SNR (13.3)
has a medium W$_{50}$ (5.7 ms),
whereas the burst with the largest W$_{50}$ (13.7 ms)
has a comparatively low SNR (6.9).
Circled dots
are the counterparts of bimodal bursts.}
\label{fig_snr_w50}
\end{figure}

Fig.\ref{fig_snr_w50} is a plot
of SNR versus W$_{50}$
for all bursts detected
from RRAT J1819$-$1458,
which shows that W$_{50}$
widely ranges from 0.5 to 13.7 ms.
The variation in SNR
that is proportional
to the peak flux density
has no apparent relation to W$_{50}$.
The widest burst with $W_{50}=13.7$ ms
has a relatively low SNR of 6.9,
whereas the brightest burst of $SNR=13.3$
has a medium W$_{50}$ of 5.7 ms.

\begin{figure}
\resizebox{\hsize}{!}{\includegraphics{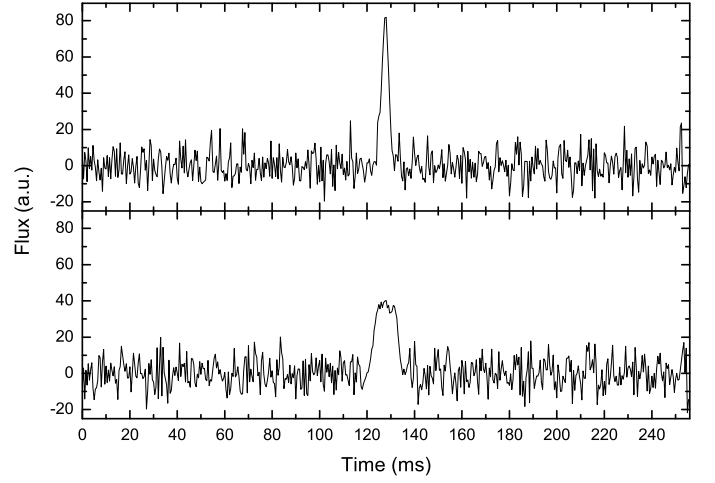}}
\caption{Profiles of two individual pulses
from RRAT J1819$-$1458
of widely different SNR and W$_{50}$.
The two bursts are
located in the middle band
of timing residuals.
SNR and W$_{50}$
of the burst in the top panel
are 11.8 and 2.9 ms
when those of the burst
in the bottom panel
are 6.7 and 10.0 ms.
The amplitude is measured
in an arbitrary unit.}
\label{fig_nar_wid}
\end{figure}

Fig.\ref{fig_nar_wid}
portrays two typical examples of
detected individual single-peak pulses
from RRAT J1819$-$1458
located in the middle band
of timing residuals
(shown in Fig.\ref{fig_int_res})
with widely different shapes.
The spiky burst
of $W_{50}=2.9$ ms
in the top panel
has a high SNR of 11.8,
whereas the broad burst
of $W_{50}=10.0$ ms in the bottom panel
has a rather low SNR of 6.7.
As shown in Figs.\ref{fig_snr_w50}
and \ref{fig_nar_wid},
profiles of detected bursts
have various shapes with
W$_{50}$ varying broadly
without any clear relation
to the peak flux density.

\section{Discussion}\label{sec_dis}
From Figs.\ref{fig_snr_w50}
and \ref{fig_nar_wid},
the profiles
of the single bursts
of RRAT J1819$-$1458
vary from pulse to pulse.
The phenomenon
is similar to the one of
that individual pulses
of a normal pulsar
differ from each other
\citep[e.g. pulses of PSR B0943+10 as reported by][]{dr99}.
Fig.\ref{fig_dist_snr_w50_res} (middle)
shows that the average SNR
of bursts
in the last band is
lower than those
in the early band.
This is consistent with
the integrated profile
from 165 individual bursts
in Fig.3 of \citet{lmk+2009}.
It is noted that
the tri-band pattern
of timing residuals
could be interpreted
in terms of a patchy radio beam
with core and conal components
\citep{lmk+2009}.
Judging from
Fig.\ref{fig_dist_snr_w50_res} (bottom),
the widths of bursts
in the last band
are prone
to be larger than
those of bursts
in the early band.
It seems that
the pulse profiles
of bursts in
the last and early bands
are different,
while all the bursts
of the two bands
are from the ``cones".
Since the spectrum
of the ``core" component
is always steeper than
those of components
from the ``cones" \citep{lm88},
a multi-frequency
and higher sensitivity study
of bursts
from one residual band
to another
may uncover more knowledge
about the phase distribution of
individual bursts.

The power-law index
of intensity CPD
for giant pulses
from PSR B1937+21
at 1420 MHz
is $\alpha=1.8$
\citep{kt00},
and for both the giant pulses
from PSR J1824$-$2452A
at 1400 MHz
and
the Crab pulsar
at 1300 MHz
is $\alpha=1.6$
\citep{kbm+06,btk2008}.
Finding a similar value of $\alpha=1.6$
also for RRAT J1819$-$1458 at 1540 MHz
suggests that
the emission mechanism
of RRAT J1819$-$1458
is perhaps similar to
that of pulsars with giant pulses.
However,
those detected giant pulses
are from pulsars
of relatively short spin periods
measured in ms
and have short pulse durations
measured in {$\mu$}s even ns,
which are quite different
from those of RRAT J1819$-$1458.
If the small amount
of bursts of low intensities
is not caused by
observation sensitivity,
the pulse intensity distribution
of RRAT J1819$-$1458
may be similar to those of
some normal pulsars
\citep{hw1974}.

Both RRAT J1819$-$1458
and PSR B0656+14
have relatively high
magnetic field strengths
and moderate characteristic ages
\citep{lmk+2009,bt1997},
and similar X-ray emission properties
are also observed for the two neutron stars
\citep{mrg+07,dcm+05}.
The intensity CPD
and emission modulation properties
of RRAT J1819$-$1458 at $\sim$1.5 GHz
resemble those of B0656+14 at 327 MHz.
Furthermore, the latter pulsar
might have been observed as
a RRAT if it had been at greater distance
\citep{wje2010,wsr+2006}.
This indicates that
RRAT J1819$-$1458 is possibly
a normal pulsar
with pulse properties
similar to PSR B0656+14,
while its quite large distance
ensures that the former is detected
as a pulsar with sporadic pulses.
This may explain
that most known RRATs
have larger DM
than most normal pulsars,
since a large DM normally corresponds to
a large distance
that can ensure that
the modulated pulses
with low flux densities
remain undetected.
However,
there is presently
only one RRAT
with $DM>500$ \dmu,
which indicates an upper limit of DM
because an excessively large distance
would ensure that
even extremely bright pulses
of a neutron star remain invisible.

\section{Conclusion}\label{sec_con}
This paper has presented
the analysis results
of 423 detected bursts
of RRAT J1819$-$1458.
A value
of $DM=195.7\pm0.3$
\dmu~with 5$\sigma$ error
is obtained.
A simple timing result is given
and the tri-band shape
of timing residuals is confirmed.
Pulses situated
in the central timing residual band
are more complicated,
including 27 bimodal bursts,
the broadest burst,
and the brightest one.
The differences
in the average SNR
and average W$_{50}$
of bursts
in the two side residual bands
are indicated.
Profiles of
four typical individual
bursts including two bimodal bursts,
and a plot of SNR versus W$_{50}$
are presented.
A possible power-law intensity CPD
with index $\alpha=1.6\pm0.2$
is measured.

Differences
in both SNR and W$_{50}$
of single pulses
in the two side bands of timing residuals
for RRAT J1819$-$1458
are briefly discussed.
Comparisons between
the flux intensity distribution
and those of giant pulses,
and,
similarities
in the emission properties
of the RRAT
and PSR B0656+14
are also briefly presented.

\begin{acknowledgements}
The authors thank
the members of Pulsar Group
at Urumqi Observatory
for the helpful discussions.
We also thank the referees
for valuable suggestions.
This work is supported
by the National Natural Science Foundation
of China
(grant 10973026, 10903019,
10935001 and 10973002),
the Natural Science Foundation
of Xinjiang Uygur Autonomous Region
of China (grant 2009211B35),
the National Basic Research Program
of China (grant 2009CB824800)
and the Knowledge Innovation Program
of the Chinese Academy of Science
(grant KJCX2-YW-T09).
\end{acknowledgements}
\bibliographystyle{aa} 
\bibliography{15953}
\end{document}